\documentclass[%
superscriptaddress,
longbibliography,
 amsmath,amssymb,
 aps, twocolumn,
 longbibliography,
]{revtex4-1}

\usepackage{graphicx}
\usepackage{dcolumn}
\usepackage{xr}
\usepackage{bm}
\usepackage[colorlinks=true,
  allcolors={blue}]{hyperref}
  \usepackage{multirow}
  \usepackage{braket}
\usepackage{iftex}

\usepackage{xcolor}

\newcommand{\fgt}{Fe$_3$GeTe$_2$}

\usepackage{changes}

\begin{document}


\title{An effective spin model on the honeycomb lattice for the description of magnetic properties in two-dimensional Fe$_3$GeTe$_2$}

\author{Georgy V. Pushkarev}
\affiliation{Theoretical Physics and Applied Mathematics Department, Ural Federal University, 620002 Yekaterinburg, Russia}

\author{Danis I. Badrtdinov}
\affiliation{Institute for Molecules and Materials, Radboud University, Heijendaalseweg 135, NL-6525 AJ Nijmegen, The Netherlands}

\author{Ilia A. Iakovlev}
\affiliation{Theoretical Physics and Applied Mathematics Department, Ural Federal University, 620002 Yekaterinburg, Russia}

\author{Vladimir V. Mazurenko}
\affiliation{Theoretical Physics and Applied Mathematics Department, Ural Federal University, 620002 Yekaterinburg, Russia}

\author{Alexander N. Rudenko}
\email{a.rudenko@science.ru.nl}
\affiliation{Institute for Molecules and Materials, Radboud University, Heijendaalseweg 135, NL-6525 AJ Nijmegen, The Netherlands}

\date{\today}

\begin{abstract} 
Fe$_3$GeTe$_2$ attracts significant attention due to technological perspectives of realizing room temperature  ferromagnetism in two-dimensional materials. Here we show that due to structural peculiarities of the Fe$_3$GeTe$_2$ monolayer, short distance between the neighboring iron atoms induces a strong exchange coupling. This strong coupling allows us to consider them as an effective cluster with a magnetic moment $\sim$5 $\mu_B$, giving rise to a simplified spin model on a bipartite honeycomb lattice with the reduced number of long-range interactions. The simplified model perfectly reproduces the results of the conventional spin model, but allows for a more tractable description of the magnetic properties of Fe$_3$GeTe$_2$, which is important, e.g., for large-scale simulations. Also, we discuss the role of biaxial strain in the stabilization of ferromagnetic ordering in Fe$_3$GeTe$_2$.          
\end{abstract}

\maketitle


\section{\label{sec1}Introduction}
The discovery of graphene in 2004 by Novoselov \emph{et al.}~\cite{Novoselov2004} attracted enormous attention to the two-dimensional (2D) materials due to their unique properties and possibility to control them by means of gate voltage, chemical doping, or strain~\cite{Geim2013, Avsar2020, Miao2021}. Since then the search and description of novel 2D materials is an actively developing research direction in material science. The discovery of 2D magnetic materials \cite{2dmag_1,2dmag_2,2dmag_3,2dmag_4} has further enhanced the interest to the field of 2D materials, opening new ways to control magnetism in two dimensions, which is prospective for applications.

One of the prominent example among 2D magnets is CrI$_3$, which was successfully exfoliated from the bulk crystal~\cite{2dmag_4} demonstrating the great advantage to control magnetic properties by electrostatic doping~\cite{CrI_1} or hydrostatic pressure \cite{CrI_2}. 
The measured Curie temperature of this material is about 45 K, slightly lower than in the bulk crystal. The other typical representative of the 2D magnet family is Fe$_3$GeTe$_2$ (FGT) \cite{Deng2018}.  The Curie temperature of FGT in the bulk phase is considerably higher $T_\mathrm{C}$ $\sim$ 220 K \cite{Deng2018, Fei2018, Verchenko2015, Bin2013}, making this material more perspective from the practical point of view. Despite a decrease of the Curie temperature down to $T_\mathrm{C}$ $\sim$ 130 K~\cite{FGT_Fei2018} in the monolayer limit, the corresponding value is still three times larger compared to CrI$_3$. Moreover, the gate control of FGT allows one to enhance $T_\mathrm{C}$ giving the opportunity to realize room-temperature ferromagnetism in 2D \cite{Deng2018}. Other studies report large anomalous Hall effect \cite{FGT_aHall}, anomalous Nernst effect \cite{FGT_aNest} and magnetic stability \cite{FGT1} making FGT a promising candidate for spintronics, caloritronics and other applications \cite{FGT3,Deng2018}.

\begin{figure}[h]
\includegraphics[width=1\columnwidth]{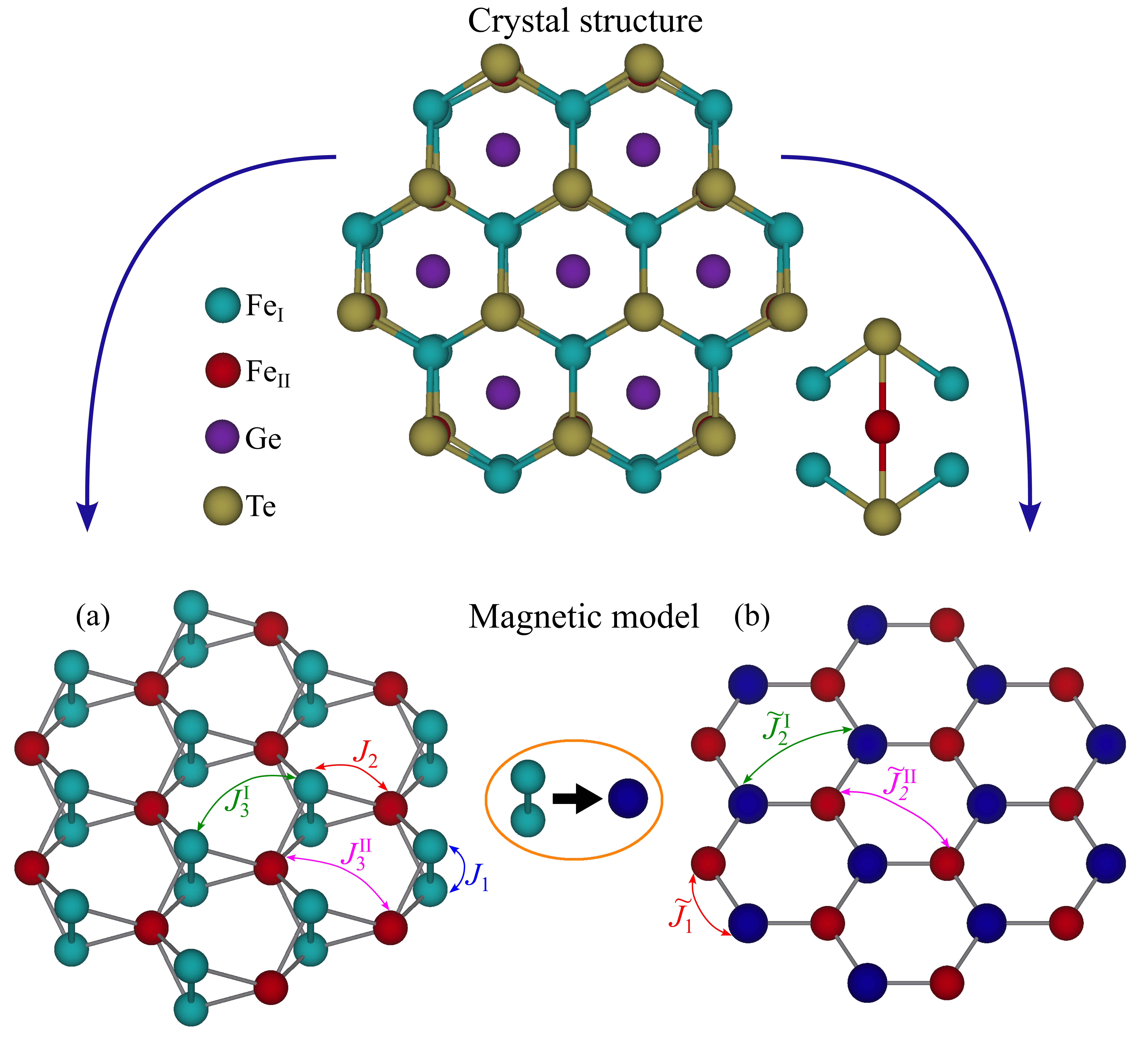}
\caption{ (Top)  Crystal structure of \fgt{} monolayer: Side and top views.  Fe$_{\rm I}$ and  Fe$_{\rm II}$ denote two inequivalent iron atoms. (Bottom) Panels (a) and (b) demonstrate the conventional and alternative magnetic  models. }
\label{fig:Structure_model}\end{figure}

Another interesting aspect of FGT is the absence of inversion symmetry \cite{Laref2020}, which gives rise to nontrivial physics and complex spin textures such as magnetic skyrmions and spin-spirals~\cite{Ding2020, Meijer2020, Park2021}, presumably emerging due to the Dzyaloshinskii-Moriya interaction (DMI)~\cite{Moriya1960}. On the other hand, some authors introduced four-spin interaction  to explain the stabilization of complex magnetic patterns~\cite{Bellaiche2022, Ado2021}. Therefore, previously proposed conventional magnetic models need to be treated carefully, considering peculiarities of the crystal structure. In particular, there are two valence types of Fe atoms in FGT: Fe$^{2+}_{\rm II}$ and Fe$^{3+}_{\rm I}$, where Fe$_{\rm I}$ atoms with comparably short distance ($\sim$2.47~\AA) form a dimer-like structure (Fig.~\ref{fig:Structure_model}). The formation of such dimers is inherent for some insulating systems~\cite{Mazurenko2014, Vasiliev2013, Riegg2014}, which challenges the description of their magnetic properties.  For instance, the short distance between the iron atoms in the dimer increases the overlap of wave functions, enhancing the hybridization and exchange interactions between the magnetic atoms. At the same time, strong interaction between the atoms leads to an ambiguity in the definition of local magnetic moments, challenging the description of magnetic properties within the spin models.
 
In this paper, we propose an alternative description for the magnetic properties of Fe$_3$GeTe$_2$ monolayer, taking advantage of its structural peculiarities. We demonstrate that the iron dimer can be considered as an effective cluster, such that the spin model of FGT reduces to a bipartite honeycomb lattice (Fig.~\ref{fig:Structure_model}). The proposed model is further justified by means of the Monte Carlo and spin dynamics simulations, which perfectly reproduce the  results of the conventional spin model in the relevant energy region at low computational cost. Additionally, we discuss the role of in-plane biaxial strain in the stabilization of magnetic order.

The rest of this paper is organized as follows. Sec.~\ref{sec:Method} briefly describes numerical methods used in this paper. In  Sec.~\ref{sec:Results}, we present our main results, including the  alternative magnetic model, and thermodynamics of FGT. The effect of biaxial strain is also discussed. In Sec.~\ref{sec:Conclusion}, we summarize our results and conclude the paper. 
 

\section{\label{sec:Method}Methods}

To study the magnetic properties of monolayer \fgt{}, we consider the following spin model:
\begin{equation}
\mathcal{H} = \sum_{i>j} J_{ij} {\bf S}_{i} {\bf S}_{j} - A\sum_{i}{S^2_{z \,i}}.
\label{eq:spin_ham}
\end{equation}
Here, $J_{ij}$  stands for the isotropic exchange interactions between iron ions, and $A$ is the single-ion anisotropy. If $A>0$, Eq.~(\ref{eq:spin_ham}) describes an easy-axis magnet with the ground-state magnetization perpendicular to the 2D plane. For simplicity reason   we map our model to unified spin $S = 1$.

Magnetic exchange interactions  were calculated using the local force theorem approach~\cite{liechtenstein1987, mazurenko2005}
\begin{equation}
 \begin{aligned}
J_{ij} = & \frac{1}{2 \pi S^2}\times& \\[5pt]
&  \times   \int  \limits_{-\infty}^{E_F} d \epsilon&\,{\rm Im} \left( \sum \limits_{m, m^{\prime},  n, n^{\prime}} \Delta^{m m^{\prime}}_i G^{m^{\prime} n}_{ij \downarrow} (\epsilon) \Delta^{n n^{\prime}}_j G^{n^{\prime} m}_{ji \uparrow} (\epsilon) \right),
 \end{aligned}
\label{eq:Exchange}
\end{equation} 
where $m, m^{\prime},  n, n^{\prime}$ are orbital quantum numbers,  $\Delta^{m m^{\prime}}_i = H^{m m^{\prime}}_{ii \uparrow} - H^{m m^{\prime}}_{ii \downarrow}$ is the intra-orbital spin-splitting energy and $\hat{G}(\epsilon)  = 1/(\epsilon - \hat{H})$ is single-particle Green's functions. In case of conventional model orbital indices run within orbitals of single site, while for conventional model they run within  orbitals of two sites (cluster). In contrast to the mapping procedure of total energies of collinear spin configurations~\cite{Xiang2011, Badrtdinov2021}, this approach allows one to estimate long-range exchange couplings without using large unit cells. We do not consider DMI in the present study, as  we  are mostly interested in the thermodynamic properties of FGT as well as in stability of the FM order, where intersite anisotropic interactions play a minor role.

Density-functional (DFT) band-structure calculations were performed within the Perdew-Burke-Ernzerhof (PBE) exchange-correlation functional~\cite{PBE} as implemented in  \emph{Vienna ab initio simulation package} ({\sc vasp}) \cite{VASP1,VASP2}. Additionally, we also performed  DFT+$U$ calculations with simplified rotationally invariant scheme~\cite{Dudarev1998} using the effective  parameter $U$ = 4  eV~\cite{He2021}, and further use these results for comparative analysis. In all these calculations, we set the energy cutoff  of the plane-wave basis to 400 eV, the energy convergence criteria to $10^{-8}$ eV and use a ($18\times18\times1$) $\Gamma$ centered grid for the Brillouin zone integration. The experimental crystal structure of bulk \fgt{} was used~\cite{Deiseroth2006}, where a vacuum space more than 16 \AA \, between monolayer replicas in the vertical $z$ direction was introduced. The positions of atoms were allowed to relax until all the residual force components on each atom were less than 10$^{-3}$ eV/\AA. From the calculated electronic structure, maximally localized Wannier functions~\cite{marzari1997} were constructed using the {\sc wannier90} package~\cite{pizzi2020} projected onto the $3d$ and $5p$ states of iron and tellurium, respectively~\cite{SM}. This ensures that the constructed Wannier functions for both spin channels are well localized, which provides an atomic-like basis for the application of Eq.~(\ref{eq:Exchange}), and for the analysis of magnetic properties.

Classical Monte Carlo (MC) simulations of the constructed spin Hamiltonian [Eq.~(\ref{eq:spin_ham})] as well as spin-wave dispersion calculations were carried out using the Uppsala Atomistic Spin Dynamics (UppASD) package~\cite{UppASD1, UppASD2}. All the results for both conventional and alternative models are obtained using the exact lattice structures presented in Fig.~\ref{fig:Structure_model}. At each temperature, we perform 20 000 MC steps for thermalization and 150 000 for measurements. Magnetization and specific heat curves are calculated for a lattice containing $90\times 90$ unit cells with periodic boundary conditions. Importantly, to obtain clear peaks in the specific heat, we averaged the graphs over ten different MC runs and additionally smoothed the resulting curve. Spin-spin correlators  $\braket{S_1 S_2}$ for nearest neighbor Fe$_{\rm I}$  and Fe$_{\rm II}$  atoms  were calculated for each considered temperatures averaging  the results over 100 MC configurations.

Spin-wave dispersion is calculated within the framework of the classical spin dynamics approach by solving the Landau-Lifshitz-Gilbert (LLG) equation.

\begin{equation}\label{LLG}
\begin{split}
\frac{d\textbf{S}_i}{dt}= -\frac{\gamma}{1+\alpha^2}\textbf{S}_i\times[-\frac{\partial H}{\partial\textbf{S}_i}+b_i(t)]-&\\[5pt] -\frac{\gamma}{|\textbf{S}_i|}\frac{\alpha}{1+\alpha^2}\textbf{S}_i\times(\textbf{S}_i\times[-\frac{\partial H}{\partial\textbf{S}_i}+b_i(t)]),
\end{split}
\end{equation}
where $\gamma$ is the gyromagnetic ratio, $\alpha$ is the damping parameter and $b_i(t)$ is a stochastic magnetic field with a Gaussian distribution arising from the thermal fluctuations. For this, we calculate the space- and time-displaced correlation functions
\begin{eqnarray*}
    C^k(r-r^\prime,t)= \langle S_r^k(t)S_{r^\prime}^k(0)\rangle - \langle S_r^k(t)\rangle\langle S_{r^\prime}^k(0)\rangle,
\label{eq_swd_c}
\end{eqnarray*}
where $\langle\dots\rangle$ denotes the ensemble average
and $k$ is the Cartesian component. The corresponding Fourier transform gives the dynamical structure factor
\begin{eqnarray*}
\notag
    \chi^k(\textbf{q}, \omega)=\frac{1}{\sqrt{2\pi}N}\times& \\[5pt]
     \sum_{r,r^\prime}e^{iq(r-r^\prime)}&\int_{-\infty}^{\infty}e^{i\omega t}C^k(r-r^\prime,t)dt,
\label{eq_swd} 
\end{eqnarray*}
with $N$ being the total number of magnetic atoms. This quantity can be probed in neutron scattering experiments of bulk systems~\cite{sqw_bulk}. The energy dispersion for spin waves presented in the simulated system is described by the positions of the peaks in $\chi^k(\textbf{q}, \omega)$~\cite{UppASD1, swd_Landau}. Thus, the resulting values at each $\textbf{q}$ vector are convoluted with a Gaussian filter and then normalized to make peaks at any $\textbf{q}$ positions visible regardless of their relative intensity~\cite{swd}.  All the results presented in this paper are obtained for a lattice of ($450\times 450$) unit cells with periodic boundary conditions.

Adiabatic magnon frequencies $\omega_{\mathbf{q} \nu}$ were calculated via diagonalizing the spin-wave Hamiltonian, defined for magnetic sublatices $\mu$ and $\nu$ in the unit cell~\cite{Rusz2005}:
\begin{equation}
\hat{\mathcal{H}}^{SW}_{\mu \nu}(\mathbf{q})  = \delta_{\mu \nu}[A \braket{S^z_\mu} + \sum_{\chi} J_{\mu \chi}(\mathbf{0})\braket{S^z_\chi} ]   - \braket{S^z_\mu} J_{\mu \nu}(\mathbf{q})  ,
\label{eq:SW}
\end{equation}
where $J_{\mu \nu}({\bf q})$ are the Fourier transform of the exchange interactions. The magnon frequencies allow us to estimate the Curie temperature within the random phase approximation (RPA) as \cite{Rusz2005,PhysRevB.102.024441}:
\begin{equation}
    T_\mathrm{C} = \frac{S}{3k_B} \left( \frac{1}{N_q} \sum_{\mathbf{q} \nu} \frac{1}{\omega_{ \mathbf{q} \nu}} \right)^{-1}. 
    \label{eq:Tc_rpa} 
\end{equation}

\section{\label{sec:Results}Results and discussion}
        
\subsection{Magnetic models}
\begin{figure}[tbp]
\includegraphics[width=0.95\columnwidth]{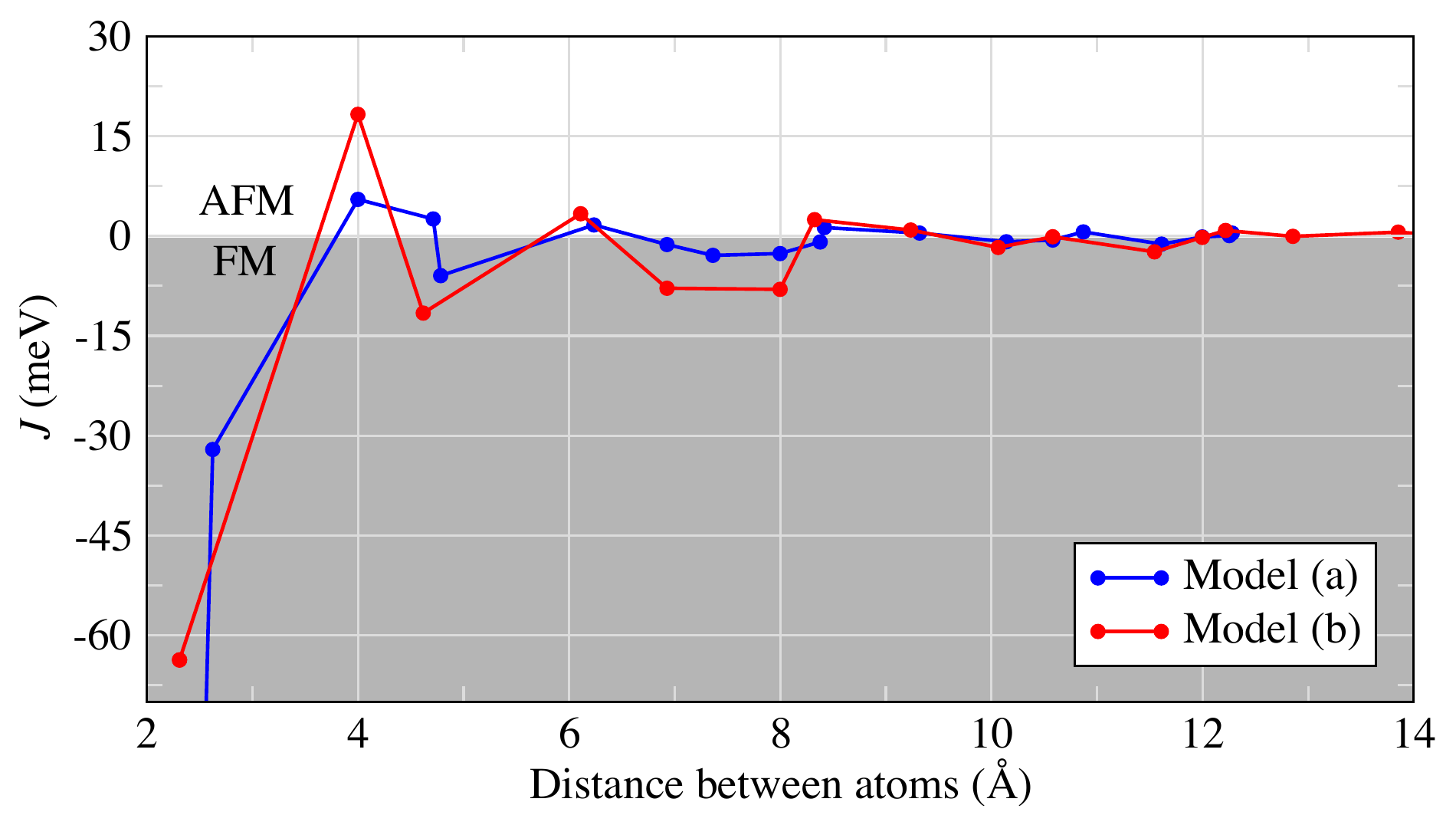}
\caption{Evolution of exchange couplings as a function of the distance between Fe$_{\rm I}$ atoms shown for the two different spin models. }
\label{fig:J_r}
\end{figure}

The resulting isotropic exchange couplings are represented in Supplemental Material~\cite{SM}. The  nearest-neighbor FM interaction $J_1$ = $-$113 meV is three times larger than  $J_2$  and an order of magnitude stronger than AFM $J_3$.  The values of these exchange couplings were additionally checked by using the total energies of collinear structures~\cite{Xiang2011, Badrtdinov2021}, which yields $J_1$ = $-$113.9 meV and $J^{\rm I}_3$ = 10.9 meV. The obtained parameters are also in reasonable agreement with recently reported values in Ref.~\onlinecite{Heinze2022}, where $J_1$ = $-$146.3 meV, $J_2$ = $-$36.2 meV and  $J_3^{\rm I}/J_3^{\rm II}$ = 10.9/4.5 meV.  Short-range interactions $J_2$ and $J_3$ originate  from the ligand-mediated superexchange  mechanism.  Due to a metallic character~\cite{Badrtdinov2023}, the exchange couplings demonstrate an oscillating RKKY-like behavior at increasing distances between the iron atoms (Fig.~\ref{fig:J_r}). These long-range interactions are essential for the simulation of magnetic systems by means of the spin models~\cite{Kashin2022}. The resulting magnetic moments in these calculations are $m_{\rm I}$ = 2.43  $\mu_B$  and $m_{\rm II}$ = 1.48 $\mu_B$.

Strong  coupling between nearest neighbour  iron  atoms allows us to propose the new model, where  two Fe$_{\rm I}$ atoms in the dimer are to be considered as an effective cluster with ten $3d$ orbitals. Within this {\it alternative} model the exchange couplings can be reformulated in terms of the interactions between these effective clusters and Fe$_{\rm II}$ atoms (Fig.~\ref{fig:Structure_model}).  The proposed alternative model, which has the form of a bipartite honeycomb lattice, is more tractable compared to the  original 3-site model as it allows to reduce the number of interactions needed for simulations, and helps to eliminate the strong coupling $J_1$ within the Fe$_{\rm I}$--Fe$_{\rm I}$ cluster. Further we will use notation of exchange  interactions for alternative model with tilde $\tilde{J}_i$. Below, we perform a systematic comparison between the magnetic properties simulated using the two different spin models, and discuss the performance of the alternative model in more detail.

\subsection{Model simulations}

\begin{figure}[tbp]
\includegraphics[width=1\columnwidth]{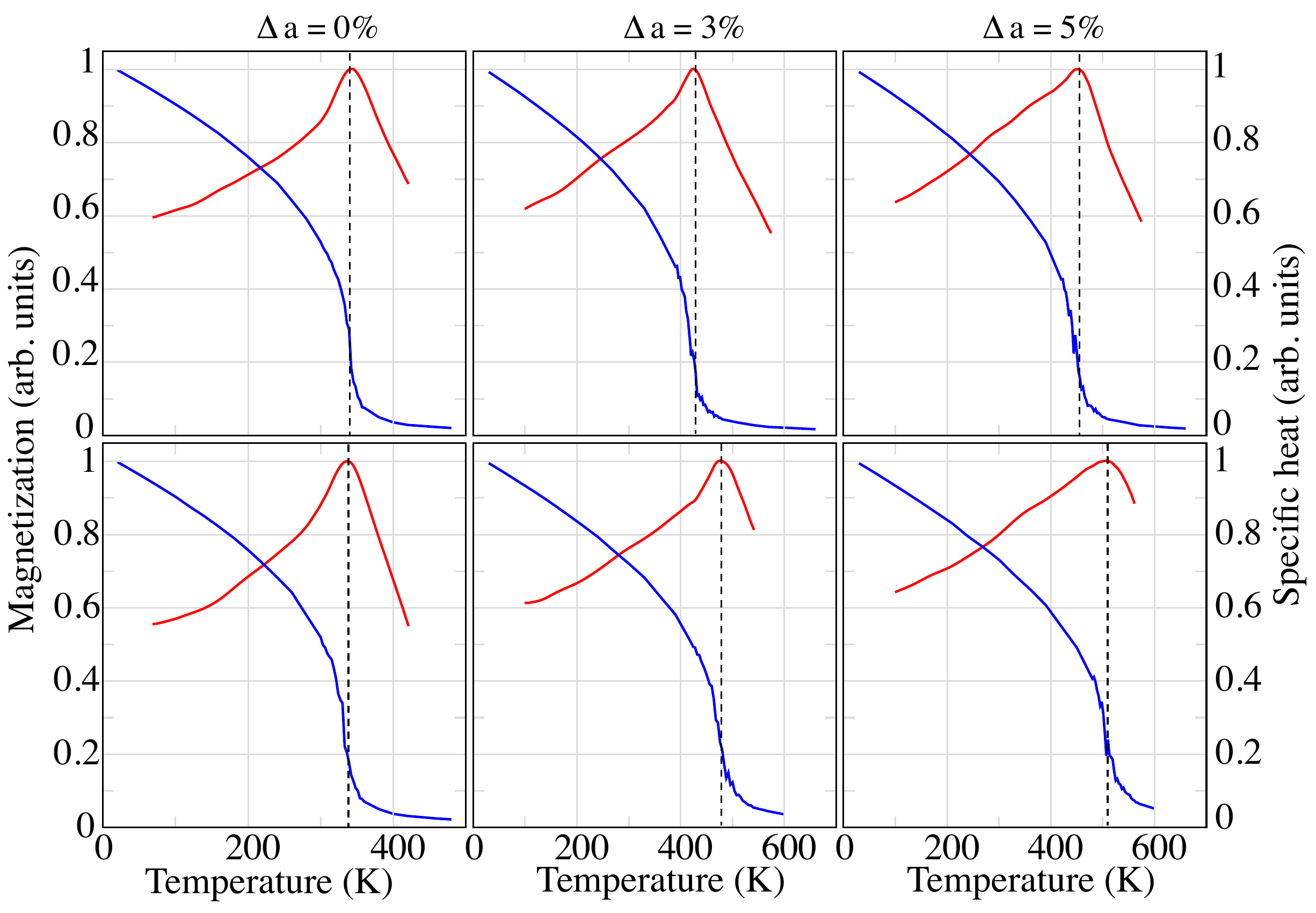}
\caption{Comparison of the magnetization (blue) and specific heat (red) curves for conventional (top) and alternative (bottom) models  under various strains.}
    \label{fig:MC_mag}
\end{figure}

Monte Carlo results for both spin models are represented in Fig.~\ref{fig:MC_mag}.  In these simulations, we use the single-ion anisotropy parameter $A$ = 0.35 meV/Fe ion, which was calculated from the total energy difference between the configurations with the magnetic moments oriented in-plane and out-of-plane directions taking spin-orbit coupling into account~\cite{He2021, Heinze2022}. Without loss of accuracy, we consider the exchange interactions around the iron atoms within the radius $r <$ 13 \AA\, (Fig.~\ref{fig:J_r}).  From Fig.~\ref{fig:MC_mag}, one can see that the calculated magnetization curves for both spin models perfectly match and drop down at $T \sim$ 340 K. At the same  time, the estimated critical temperature via the adiabatic magnon spectra  gives  $T_\mathrm{C}$ = 336 K and 328 K for the conventional and alternative models, respectively.

 \begin{figure}[tbp]
 \includegraphics[width=1\columnwidth]{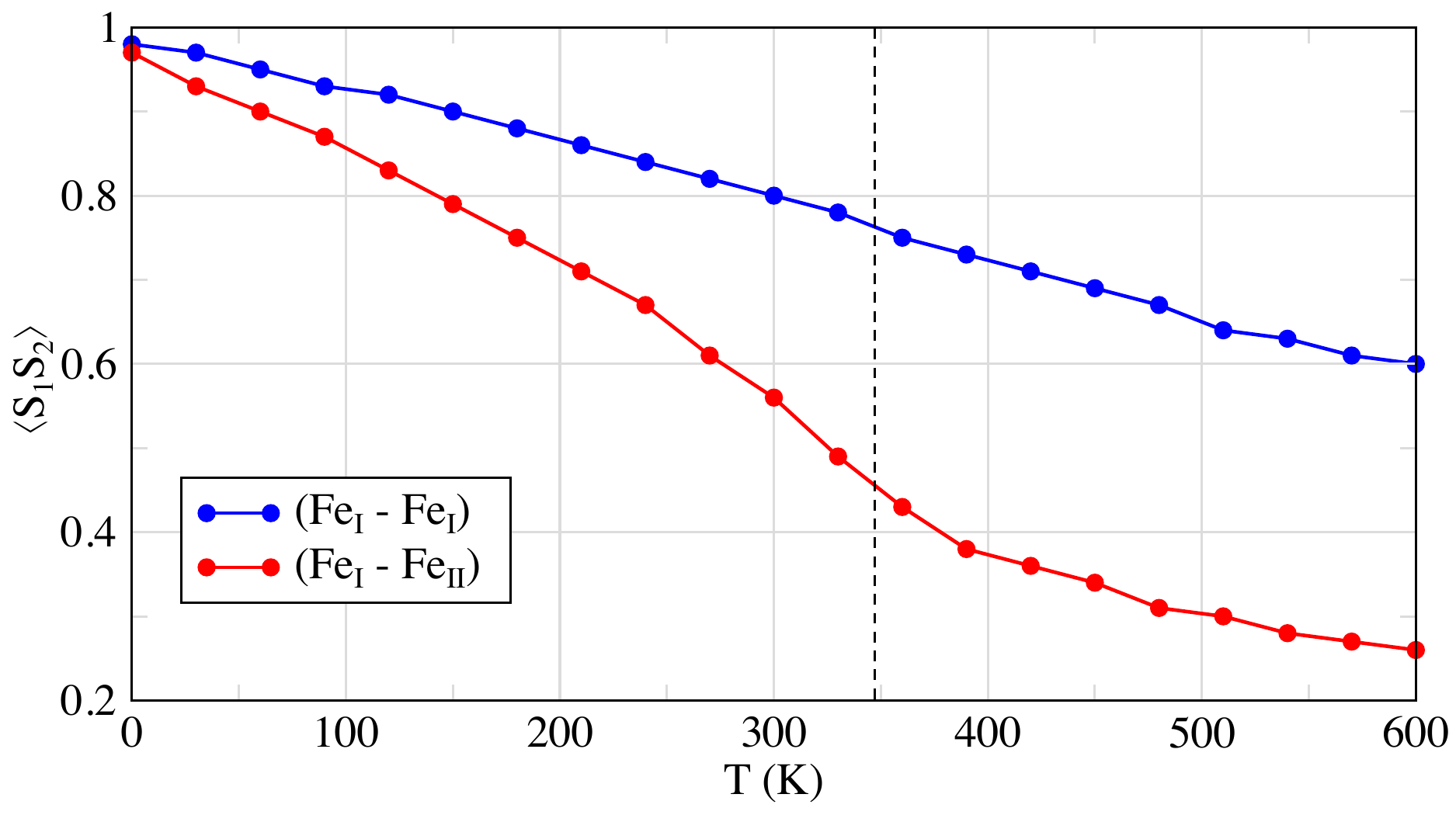}
 \caption{Temperature dependence of  spin-spin correlator $\braket{S_1 S_2}$ calculated between nearest neighbor Fe$_{\rm I}$ -- Fe$_{\rm I}$  and Fe$_{\rm I}$ -- Fe$_{\rm II}$  atomic pairs (see Fig.~\ref{fig:Structure_model}). The dashed line denotes the critical temperature.} 
     \label{fig:SRMO}
 \end{figure}
 
\begin{figure}[tbp]
 \includegraphics[width=1\columnwidth]{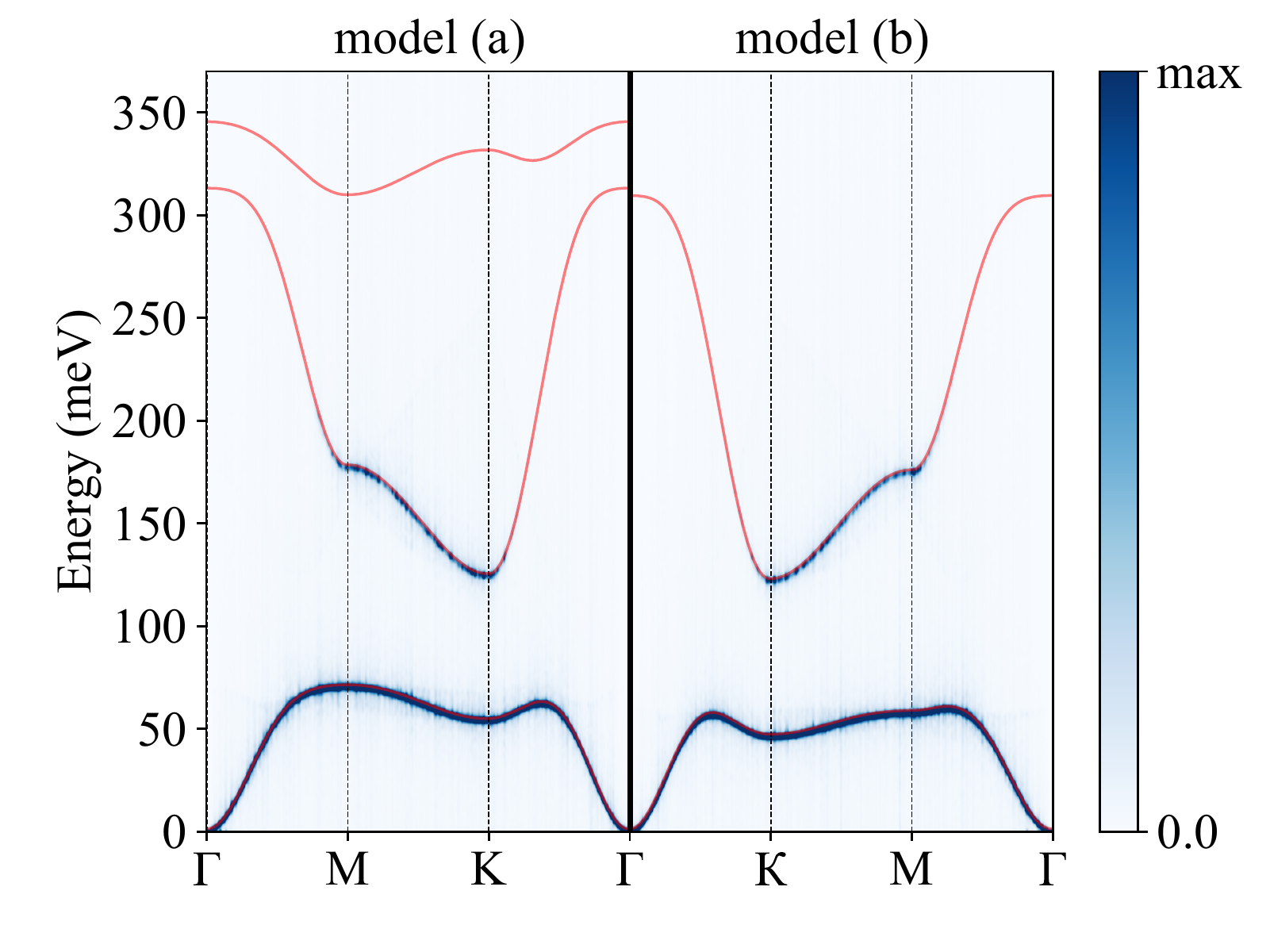}
 \caption{Comparison of the magnon spectral function intensity map for conventional model (a) (left) and alternative model (b) (right), obtained via spin dynamics calculations at $T = 5$~K and damping constant $\alpha=10^{-3}$. Red lines correspond to the adiabatic magnon spectra.}
     \label{fig:Spin_dynamics}
 \end{figure}

Figure~\ref{fig:Spin_dynamics} shows the adiabatic magnon dispersion relation calculated for the two models. The resulting low energy (acoustic and optical) magnon branches are in good agreement between the models. The spin-wave stiffness constants are also close to each other and equal to $D \approx 445$ meV$\cdot$\AA$^2$ and $D \approx 408$ meV$\cdot$\AA$^2$ for the conventional and alternative models, respectively. The magnon gap for both models is 1.36 meV since it is independent of the exchange interactions. The most prominent feature in the magnon spectrum of the conventional model is the presence of a nearly flat branch at high energies, which is mainly originated from the nearest-neighbor $J_1$ interaction between iron atoms. In the alternative model, the flat branch is absent (Fig.~\ref{fig:Spin_dynamics}) because the two Fe$_{\rm I}$ atoms are replaced by a single cluster. The energies of the corresponding flat branch are extremely high for magnons. This suggests that these states are strongly damped by other spin (or even electron) excitations in the system.

To examine whether one can expect to observe this flat branch in the experiments, we calculate the magnon spectral function for the excited spin waves using spin dynamics simulations, which is a theoretical analog of the inelastic neutron scattering. Our calculations do not take coupling with the Stoner excitations into account, but allows us to elucidate on the magnon damping resulting from the localized spin excitations. As can be seen from Fig.~\ref{fig:Spin_dynamics}, both models behave in a similar way, reproducing the adiabatic magnon dispersion with decaying intensity toward high energies. Even for the conventional model, the high energy flat branch is strongly damped. The obtained behavior is not  unusual. For example, the authors of Ref.~\onlinecite{Taroni2011} demonstrate that the intensities of the optical modes near the center of the Brillouin zone are suppressed because of the specific dynamical properties of the Heisenberg model, leading to strong damping of the standing modes observed in the adiabatic spin-wave calculations. We suppose that the absence of the flat branch in the spin dynamics simulation of \fgt{}~is in line with this scenario. Another reason why high-energy optical mode is absent may be associated with the strong short range magnetic order between Fe$_{\rm I}$ atoms, which is demonstrated by the corresponding spin-spin correlation functions in Fig.~\ref{fig:SRMO}. Strong interaction within the Fe$_{\rm I}$--Fe$_{\rm I}$ pair suppresses excitations of individual spins.

The obtained Curie temperature overestimates the  experimental  $T_\mathrm{C}$ $\sim$ 200 K \cite{Deng2018, Fei2018, Verchenko2015, Bin2013}, which is likely due to overestimation of the exchange interactions. For  $3d$ metals it is instructive to apply a Hubbard $U$ correction within DFT+$U$ method. However, our results show that this method  results in even further overestimation of exchange interactions~\cite{SM} and  $T_\mathrm{C}$. This discrepancy could be attributed to a limited applicability of the localized spin models to FGT, which neglect coupling to the electronic subsystem, especially important for metals. 
Partially itinerant character of FGT is also suggested by the non-integer magnetic moments on Fe atoms.
Previous works demonstrated a significant renormalization of the electron spectrum~\cite{FGT_aHall} as well as site-dependent correlations \cite{Kim2022}, as follows from dynamical mean-field studies. An explicit consideration of the electron subsystem as well as electron correlation effects indeed contribute to the reduction of the Curie temperature in \fgt{}, as it  has been recently demonstrated \cite{Ghosh2023}. At the same time, the description of magnetism within the localized spin models still remain possible with renormalized exchange interactions. This, therefore, is not expected to limit the applicability of the alternative model we propose in our study.


\begin{table}[h!]
\centering
\caption {Comparison of the Curie temperature $T_\mathrm{C}$ (in K) calculated for \fgt{} via MC simulations and RPA approach for the conventional and alternative magnetic models of \fgt{}~under strain. Values in brackets correspond to results of simulations using DFT+$U$  parameters with $U$ = 4~eV.} 
\begin{ruledtabular}
\begin {tabular}{c|cc|cc}
  &  \multicolumn{2}{c}{Conventional model (a)} & \multicolumn{2}{c}{Alternative model (b) }  \\
 $\Delta a$ &     MC   &   RPA  &    MC  &  RPA   \\ 
 \hline
0           &    $\sim341$ (631)  &   336 (544)  &    $\sim337$ (685) &  328  (618)   \\
3\%         &    $\sim424$   &   403  &    $\sim477$ &  452   \\ 
5\%         &    $\sim451$   &   407  &    $\sim507$ &  445   \\
\end {tabular}
\end{ruledtabular}
\label{tab:Tc}
\end {table}

\subsection{Effect of biaxial strain}

The constructed models allow us to study  the magnetic properties of FGT under in-plane biaxial strain. For this purpose, we vary the lattice constant  $a$ from $-$5\% (compressive) to 5\% (tensile) and recalculated the isotropic exchange couplings. The resulting magnetic  moments of Fe$_{\rm II}$ remain nearly the same (1.50 $\mu_B$), while the moments of Fe$_{\rm I}$ increase up to 2.6 $\mu_B$ in case of the tensile strain, and reduce down to 1.53 $\mu_B$ under compression. This is in agreement with the results reported previously~\cite{Sun2020}, also showing that tensile strain significantly enhances the FM stability and increases $T_\mathrm{C}$~\cite{Zhu2021}.

\begin{figure}[tbp]
\includegraphics[width=0.9\columnwidth]{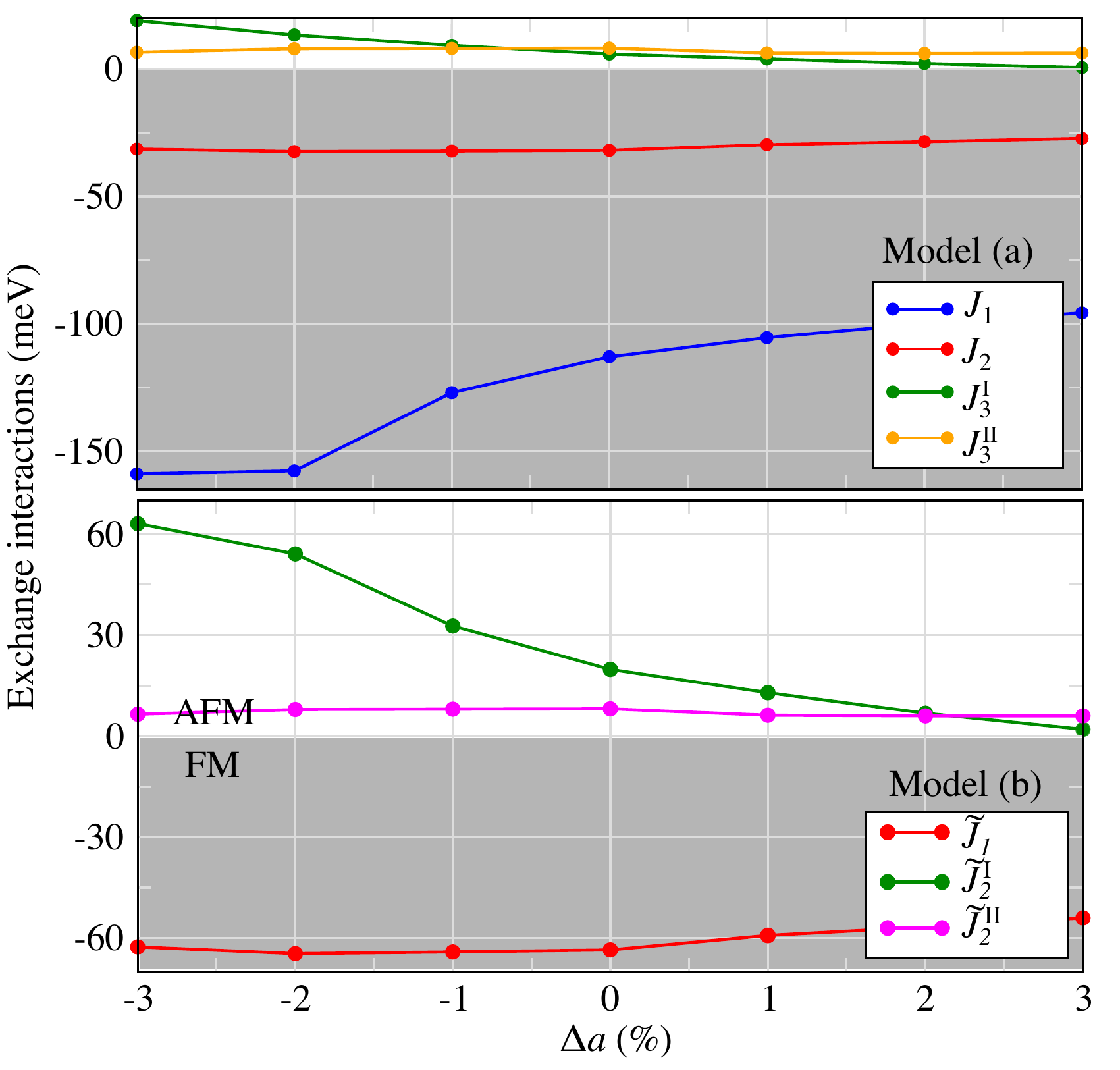}
\caption{Evolution of the nearest-neighbor and next-nearest-neighbor exchange interactions in the conventional [Fig.~\ref{fig:Structure_model}(a)] and alternative [Fig.~\ref{fig:Structure_model}(b)] spin models for various values of strain. For details see the crystal structure,  corresponding exchange interactions are given in the same color scheme. Note that short distance interaction between Fe$_{\rm I}$ atoms ($J_1$ shown with blue solid line) in Model (a) is absent in Model (b).   
} 
    \label{fig:delta_J}
\end{figure}

\begin{figure*}[th]
\includegraphics[width=1\textwidth]{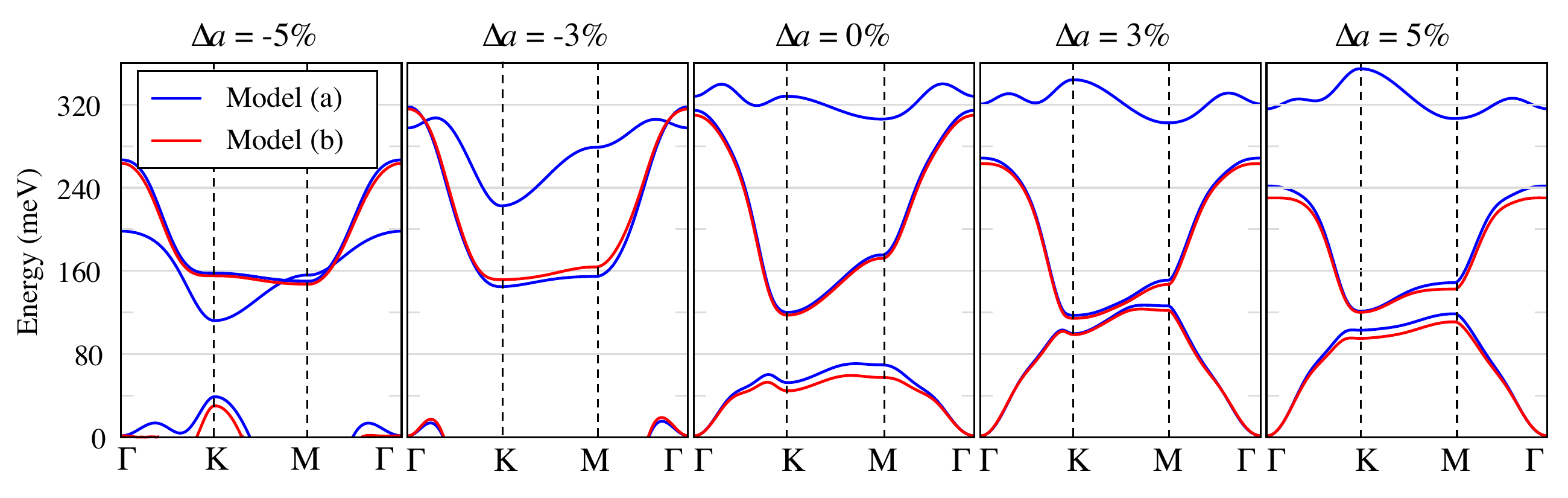}
\caption{Evolution of the adiabatic magnon spectra calculated for the conventional [Fig.~\ref{fig:Structure_model}(a)] and alternative [Fig.~\ref{fig:Structure_model}(b)] spin models of \fgt{}~under strain.} 
    \label{fig:ams}
\end{figure*}

For the conventional model, the nearest-neighbor FM interaction $J_1$ increases its absolute value under pressure  (Fig.~\ref{fig:delta_J}), which, however, does not affect the Curie temperature keeping in mind the adiabatic magnon spectra. The next-nearest-neighbor FM interaction $J_{2}$ demonstrates only a moderate change under strain. More importantly, the third-nearest-neighbor interaction between Fe$_{\rm I}$ atoms, i.e. $J_3^{\rm I}$ changes its sign  from AFM to FM at $\Delta a \approx 3$\%. A similar tendency can be seen for $\tilde{J}_2^{\rm I}$ within the alternative model in Fig.~\ref{fig:delta_J}(b). Such a behavior reduces magnetic frustration, making the FM order more preferable under tensile strain. This is further demonstrated by the Monte Carlo simulations as well as by calculations within RPA, both leading to an increase of the Curie temperature at $\Delta a > 0$ (see Fig.~\ref{fig:J_r} and Table~\ref{tab:Tc}). 
 
At the same time, compressive strain suggests a destabilization of the FM order due to the presence of AFM exchange interactions. Indeed, from the adiabatic magnon spectra (Fig.~\ref{fig:ams}) one can see that the acoustic mode becomes imaginary at finite wave vectors, indicating instability of the FM ground state in compressed \fgt{}. 

\section{\label{sec:Conclusion}Conclusion}

In this work, we present a systematic description of the magnetic properties of monolayer Fe$_3$GeTe$_2$ using a combination of spin Hamiltonians and $ab$ $initio$ calculations. A strong coupling between the nearest iron atoms motivates us to consider them as an effective cluster with a magnetic moment of $\sim$5 $\mu_B$, giving rise to a simplified spin model on a bipartite honeycomb lattice.  This lattice is in many respects a more tractable and fundamentally important model in physics, which permits analytical calculations and might be important for the development or extensions of many-body theories related to collective excitations in 2D honeycomb magnets~\cite{Ghosh2022}. The alternative spin model perfectly reproduces the results of the conventional three-site model in the moderate-energy region, which is demonstrated by simulating the magnon spectra as well as thermodynamical properties. The alternative model allows us to reduce the number of long-range interactions needed for modeling of metallic systems, which is important, e.g., for large-scale simulations. We also find that the stability of the FM ordering in monolayer \fgt{}~can be enhanced under tensile strain.

\begin{acknowledgements}
This work was supported by the Russian Science Foundation, Grant No. 21-72-10136.
\end{acknowledgements}


%

\pagebreak
\widetext
\begin{center}
\textbf{\large Supplemental Material: An effective spin model on the honeycomb lattice for the description of magnetic properties in two-dimensional Fe$_3$GeTe$_2$}
\end{center}
\setcounter{equation}{0}
\setcounter{figure}{0}
\setcounter{table}{0}
\setcounter{page}{1}
\makeatletter
\renewcommand{\theequation}{S\arabic{equation}}
\renewcommand{\thefigure}{S\arabic{figure}}
\renewcommand{\bibnumfmt}[1]{[S#1]}
\renewcommand{\citenumfont}[1]{S#1}

\section{\label{si:bands}Band structures and  Wannier functions fit }
\begin{figure}[!h]
\includegraphics[width=1\columnwidth]{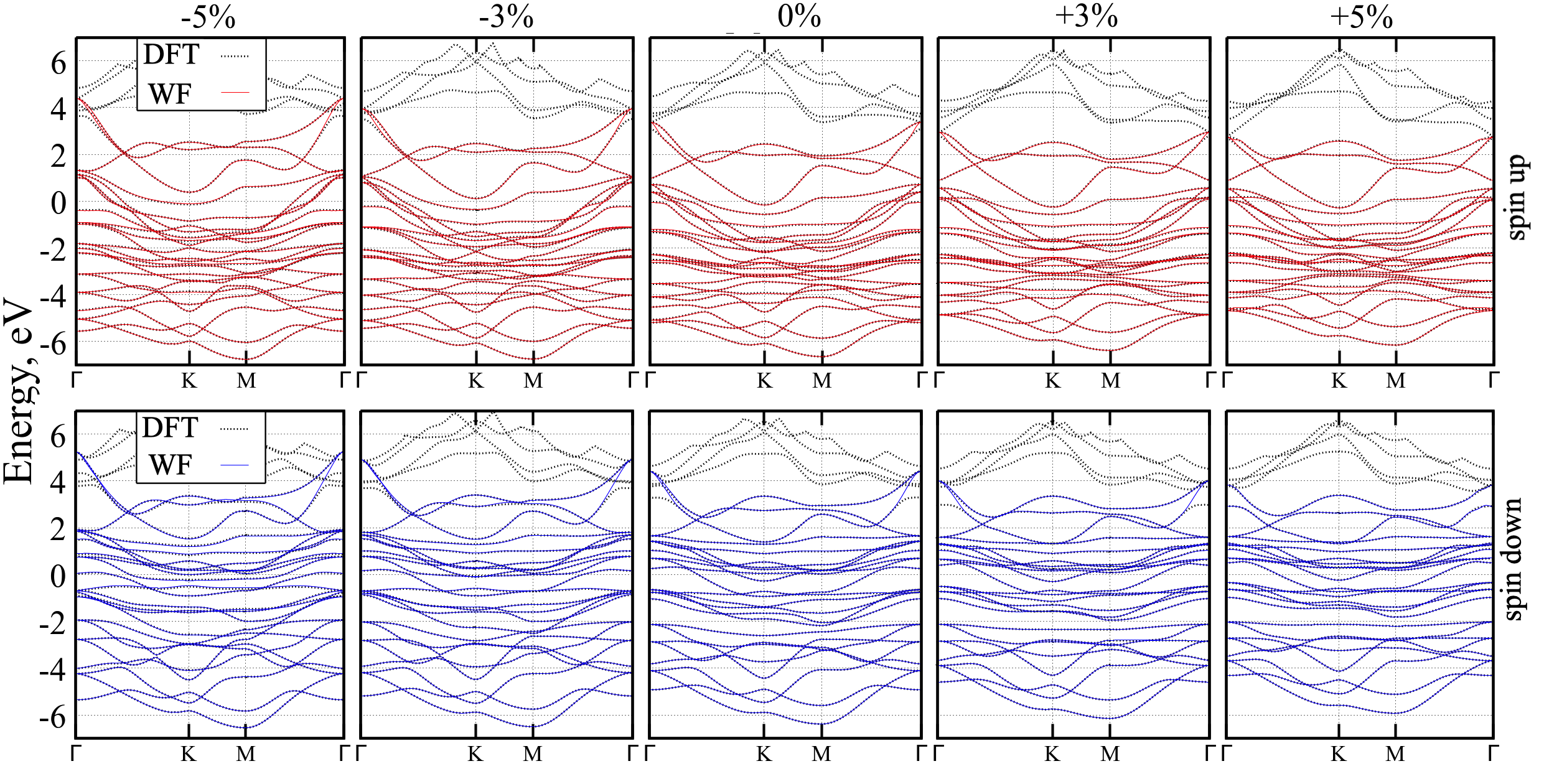}
\caption{Evolution of band structures for various values of strain. Solid and dashed lines represent DFT and the result of  Wannier functions  fit.}
    \label{figS1}
\end{figure}

In Fig.~\ref{figS1} we show spin resolved electronic band structure for all considered in-plane biaxial strain values.  

\pagebreak

\section{\label{si:ISO}Isotropic exchange interactions}

In Table~\ref{tab:Exchange_couplings} we present the isotropic exchange interactions between iron  atoms (or clusters) in \fgt{}, calculated at ambient pressure. 

\begin{table}[!h]
\centering
\caption { Main exchange  coupling (in meV) of \fgt{} within  conventional  and  alternative magnetic model calculated using  DFT and DFT+$U$ ($U = 4$ eV) electronic structure.  }
\begin{ruledtabular}
\begin {tabular}{ccccc|ccccc}
   \multicolumn{5}{c|}{ Conventional  model } & \multicolumn{5}{c}{ Alternative   model } \\
 \hline
 $i$  &   $d(\AA)$ & $\mathrm{Fe_i - Fe_j}$  &  $J^{\rm DFT}_{i}$ & $J^{\rm DFT+U}_{i}$  &  $i$  &   $d(\AA)$  & $\mathrm{Fe_i - Fe_j}$  &  $\tilde{J}^{\rm DFT}_{i}$    & $\tilde{J}^{\rm DFT+U}_{i}$   \\
 \hline 
1 &  2.493 & I - I   &  -113.0    &  -112.4   &  \multirow{2}{*}{1} &  \multirow{2}{*}{2.309} & \multirow{2}{*}{I - II} &   \multirow{2}{*}{-63.5 }  & \multirow{2}{*}{-112.9}  \\
2 &  2.624 & I - II  &  -32.0     &  -56.8     &     &      &             \\
3 &  3.991 & (I - I)/(II - II)    &  5.9/8.1     &  6.0/7.2  &  2  & 3.991  &  (I - I)/(II - II)  &    19.8/9.1   & 7.0/7.3       \\
4 &  4.712 & I - I   &  3.0       &  -3.0     &  \multirow{2}{*}{3} &  \multirow{2}{*}{4.617} & \multirow{2}{*}{I - II}   &   \multirow{2}{*}{-11.7} &  \multirow{2}{*}{-24.0}    \\
5 &  4.782 & I - II  &  -6.0      &  -12.1    &     &        &                \\
6 &  6.234 & I - II  &  1.7       &   3.4     &  4  & 6.108  & I - II &  3.4    &  7.1      \\ 
7 &  6.926 & (I - I)/(II - II)   &  -1.7/-4.0 &    -2.8/-0.1 & \multirow{2}{*}{5} &  \multirow{2}{*}{6.926} & \multirow{2}{*}{(I - I)/(II - II)} &  \multirow{2}{*}{-9.1/-4.0} & \multirow{2}{*}{0.7/0.0} \\
8 &  7.361 & I - I   &  -3.2     &  3.0      &     &        &               \\
9 &  7.982 & (I - I)/(II - II)    &  -2.3/1.2   & -3.7/-0.8 &  6  & 7.982  & (I - I)/(II - II) &  -7.3/1.2 & -7.4/-0.8    \\
10& 8.377  & I - I    &  -0.9     &  0.0      &\multirow{2}{*}{7} &  \multirow{2}{*}{8.324} & \multirow{2}{*}{I - II} &   \multirow{2}{*}{2.4} & \multirow{2}{*}{-0.4}  \\ 
11& 8.417  & I - II   &  1.2      & -0.2      &        &       &                \\
12& 9.319  & I - II   &  0.3      &  0.2      &  8     &   9.235  & I - II   &   0.5   &  0.4         \\
13& 10.141 & I - II   &  -0.9     &  0.0      &  9     &   10.063 & I - II   &  -1.8   &  0.0         \\
14& 10.580 & I - I    &  -0.3/0.4  & 0.8/-0.2  &  10    &   10.580 & (I - I)/(II - II) &  0.3/0.4  & -0.3/-0.2    \\
15& 10.870 & I - I    &  0.5       & -0.9      &\multirow{2}{*}{11} &  \multirow{2}{*}{11.544} & \multirow{2}{*}{I - II}  &   \multirow{2}{*}{-2.6} &  \multirow{2}{*}{1.4}    \\  
16& 11.611 & I - II   &  -1.3      &  0.7      &        &           &           \\
17& 11.973 & I - I    &  -0.3/0.2  &  0.1/-0.2 &  12    &   11.973  & (I - I)/(II - II) &  -0.8/0.2   & -1.5/-0.2   \\ 
18& 12.253 & I - I    &  -0.1      &  -0.7     &\multirow{2}{*}{13} &  \multirow{2}{*}{12.217} & \multirow{2}{*}{I - II}  &  \multirow{2}{*}{0.8} & \multirow{2}{*}{-0.1}    \\
19& 12.280 & I - II   &  0.4       &   0.0     &        &           &           \\ 
20& 12.915 & I - II   &   0.0      &   0.0     &  14    &  12.854  & I - II  &   0.0   &  0.0   \\
  \end {tabular}
\end{ruledtabular}
\label{tab:Exchange_couplings}
\end {table}

\end{document}